\documentclass[reprint]{revtex4-2}

\usepackage{graphicx}
\usepackage{dcolumn}
\usepackage{bm}
\usepackage{amsmath}
\usepackage{color}
\usepackage{overpic}
\usepackage{hyperref}

\begin{document}

\preprint{APS/123-QED}

\title{Upper Bound on the Cosmic Baryon Chemical Potential from Lepton-Flavor Asymmetry}
\author{Francesco Di Clemente}
\email{fdicleme@central.uh.edu}
\affiliation{Department of Physics, University of Houston, Houston, TX 77204, USA }

\author{Alessandro Drago}
\affiliation{INFN, Sezione di Ferrara, via Saragat 1, I-44122 Ferrara, Italy}
\affiliation{Department of Physics and Earth Science, University of Ferrara, via Saragat 1, I-44122 Ferrara, Italy}
\author{Lorenzo Formaggio}
\affiliation{Department of Physics, University of Houston, Houston, TX 77204, USA }
\author{Claudia Ratti}
\affiliation{Department of Physics, University of Houston, Houston, TX 77204, USA }
\author{Volodymyr Vovchenko}
\affiliation{Department of Physics, University of Houston, Houston, TX 77204, USA }
\author{Geetika Yadav}
\affiliation{Department of Physics, University of Houston, Houston, TX 77204, USA }
\date{\today}

\begin{abstract}
We study the early Universe trajectory around the QCD transition in lepton-flavor-asymmetric cases with small total lepton asymmetry ($|\ell|\lesssim 10^{-2}$), 
while allowing large individual lepton asymmetries.
For each temperature, we find an upper bound on the baryon chemical potential $\mu_{\mathrm B}(T)$: $\tau$--$\mu$ asymmetric cases exhibit a local maximum, whereas $\mu$--$e$ cases approach a limiting curve. Thus, even extreme lepton-flavor asymmetry alone cannot reach a first-order region, unless the critical point is moved to a substantially lower $\mu_\mathrm{B}/T$ because of the nonzero $\mu_\mathrm{Q}$.
Therefore, we constrain the QCD-era relic to the standard scenario of a chiral crossover transition.
\end{abstract}

\maketitle

\textit{Introduction.}---
We describe the early Universe evolution as a trajectory in the temperature-chemical potential space. In particular, we focus on its projection on the $T$-$\mu_\mathrm{B}$ plane, with $T$ the temperature and $\mu_\mathrm{B}$ the baryon chemical potential. In the standard picture, during the QCD transition, the baryon asymmetry $\eta _\mathrm{B}$ and the lepton asymmetry $\ell$ are both tiny $\big(\mathcal{O}(10^{-11})\big)$, so the trajectory remains near $\mu_\mathrm{B}\simeq 0$ and passes through the QCD crossover \cite{Borsanyi:2020fev}. 
It has been argued that large, flavor-asymmetric lepton asymmetries -- e.g., in the $\mu$ or $\tau$ sector -- could shift the trajectory to higher $\mu_\mathrm{B}$ and $\mu_\mathrm{Q}$ values \cite{PhysRevLett.121.201302,Zheng2025}.
This could lead to the formation of a pion condensate during cosmic evolution \cite{Vovchenko:2020crk,PhysRevD.105.123533}.
Another possibility is crossing a first-order chiral QCD transition at finite baryon density \cite{PhysRevLett.128.131301,Domcke:2025lzg}. 
The idea of a cosmological first-order phase transition is of great interest, as it can lead to relics such as density inhomogeneities, dark matter in the form of low-mass primordial black-holes or strangelets, and a stochastic gravitational-wave background \cite{Witten:1984rs,DiClemente:2024lzi,Musco:2023dak,Gonin:2025uvc,Shao:2024ygm,NANOGrav:2023gor,Gouttenoire:2023bqy}.

In this Letter, we re-examine this idea by calculating the trajectory, adopting both a free quark-gluon plasma (QGP) and a state-of-the-art lattice QCD-based EoS \cite{Abuali:2025tbd} for $2+1+1$ flavor QCD.
This type of framework has been considered in Refs. \cite{PhysRevLett.121.201302,PhysRevD.105.123533,PhysRevLett.128.131301,Formaggio:2025nde}, which interpolates between the Hadron Resonance Gas (HRG) model at low temperatures ($T \lesssim 140\,\mathrm{MeV}$) and an interacting 4-flavor QGP at high temperatures. 
We also check the results within the HRG model with pion interactions from \cite{Vovchenko:2020crk}, implemented in Thermal-FIST package \cite{Vovchenko:2019pjl}.
We assume thermal and chemical equilibrium, $\beta$-equilibrium for weak processes, and charge neutrality. We allow flavor-dependent lepton asymmetries $(\ell_e,\ell_\mu,\ell_\tau)$ while keeping the \emph{total} lepton asymmetry $\ell=\ell_e+\ell_\mu+\ell_\tau$ within cosmological bounds $\big(|\ell|\lesssim 10^{-2}\big)$ \cite{SergioPastor_2010, Barenboim_2017, Oldengott_2017,Lattanzi:2024hnq}. The trajectory $(T,\mu_\mathrm{B})$ then follows from the neutrality constraints and $\beta$-equilibrium relations among chemical potentials.
Although the lepton asymmetry $\ell$ is constrained at late times, the individual lepton asymmetries can partially cancel depending on the cosmological scenario, e.g., $\ell_e=0$, $\ell_\mu=-\ell_\tau$ (electro-phobic) or $\ell_\tau=0$, $\ell_e=-\ell_\mu$ ($\tau$-phobic) \cite{Castorina:2012md, Mangano:2011ip}. In such flavor-asymmetric cases, we observe a particular behavior that is worth considering.

Our main result is simple: \emph{the presence of flavor-asymmetric lepton asymmetries does not push the cosmological trajectory to high enough baryon densities to undergo a first-order QCD transition}. The key is a robust inversion in $\mu_\mathrm{B}(T)$ enforced by conservation laws, $\beta$-equilibrium, and the lepton-mass hierarchy. This mechanism depends only weakly on the details of the equation of state: interactions shift quantitative values (e.g., the maximum $\mu_\mathrm{B}$ reached), but they do not remove the upper bound on $\mu_\mathrm{B}$ along trajectories with $|\ell|\ll 1$. Therefore, unless the first-order region extends to $\mu_\mathrm{B}/T\sim1$, which is disfavored by available lattice QCD constraints on the QCD critical point, flavor-asymmetric patterns with standard small $|\ell|$ are unlikely to cause the Universe cross it. The results of this Letter motivate a focus on non-standard early-Universe scenarios \cite{Affleck:1984fy,Gao:2024fhm} if one aims to explain primordial black holes or a stochastic gravitational-wave background via a first-order QCD transition.

\textit{Asymmetric case $\ell_e \sim 0$.}---
We first consider the electrophobic case, corresponding to the muon-tauon asymmetric scenario ($\ell_\mu=-\ell_\tau$ and $\ell_e\sim0$).
We start our analysis of the limiting behavior in the $\mu_\mathrm{B}$ direction by using the free QGP EoS, where the results can be obtained in more explicit form.

At temperatures well above the QCD transition (i.e., at the initial point of our calculation, $T\sim300$  MeV), both $\tau$ and $\mu$ populations are similarly abundant due to large corresponding lepton flavor asymmetries.
These abundances are regulated by large and opposite charged-lepton chemical potentials, $\mu_\tau$ and $\mu_\mu$.
The charge neutrality condition is dominated by the large $\mu$ and $\tau$ populations,
\begin{equation}
Q_\mathrm{lep}(T)\simeq\big[n_{\tau^+}-n_{\tau^-}\big]+\big[n_{\mu^+}-n_{\mu^-}\big]\,,
\end{equation}
while the role of quarks (baryon sector) is small, and so are the values of $\mu_\mathrm{B}$.
At these temperatures ($T \gtrsim 200$ MeV), a trajectory with larger $\ell_\tau$ typically lies to the left (smaller $\mu_\mathrm{B}$) compared to one with a smaller $\ell_\tau$. The reason is that for larger $\ell_\tau$ the cancellation of the charged lepton densities $n_{\tau^-}\approx n_{\mu^+}$ is more effective, so charge neutrality is essentially handled by leptons and the quark sector does not need to raise $\mu_\mathrm{B}$. In contrast, if the $\ell_\tau$ population is smaller, $\tau^-$ is suppressed earlier while $\mu^+$ remains relatively abundant; hence, neutrality must be restored by the quark sector, which pushes $\mu_\mathrm{B}$ up.

As the temperature decreases to intermediate values, the heavy $\tau^-$ are suppressed much earlier than the lighter $\mu^+$ because of their larger mass (with the temperature depending on how large $\ell_\tau$ is).
At some point, the tauon density becomes negligible relative to that of $\mu^+$.
Therefore, to keep the system charge-neutral, the quark sector must supply a compensating negative charge.
The combined adjustment produces a rise of $\mu_\mathrm{B}(T)$ as the temperature is lowered.

An effective criterion for when $\tau^-$ becomes negligible can be found by comparing the $\tau$ chemical potential to its mass:
\begin{equation}
\mu_\tau(T)\simeq  m_\tau\,.
\end{equation}
Increasing $|\mu_\tau|$ through the lepton asymmetry delays this condition to smaller $T$, so that the onset of the charge imbalance (and thus the $\mu_\mathrm{B}$, and correspondingly, the $\mu_\mathrm{Q}$ peaks occur at a lower temperature. Indeed, for large flavor asymmetry, the charged-lepton densities remain relatively close to each other over a broader temperature window around the QCD scale.

If, instead, $\ell_{\tau}$ is smaller, both $\tau$ and $\mu$ number densities are reduced at the same initial $T$, but the lighter muons remain thermally relevant down to lower temperatures. The leptonic charge is then less efficiently self-canceled ($n_{\tau^-}\ll n_{\mu^+}$), so electric neutrality cannot be satisfied by leptons alone. The quark sector must provide a negative charge earlier, which shifts the solution towards larger $\mu_\mathrm{B}$. Nevertheless, since $\ell_\mu=-\ell_\tau$, the total lepton asymmetry remains small, and $\mu_\mathrm{B}$ increases only up to a maximum value and never reaches arbitrarily large values.

\begin{figure}
\begin{centering}
\hspace{-1cm}

\begin{overpic}[width=1\linewidth,keepaspectratio]{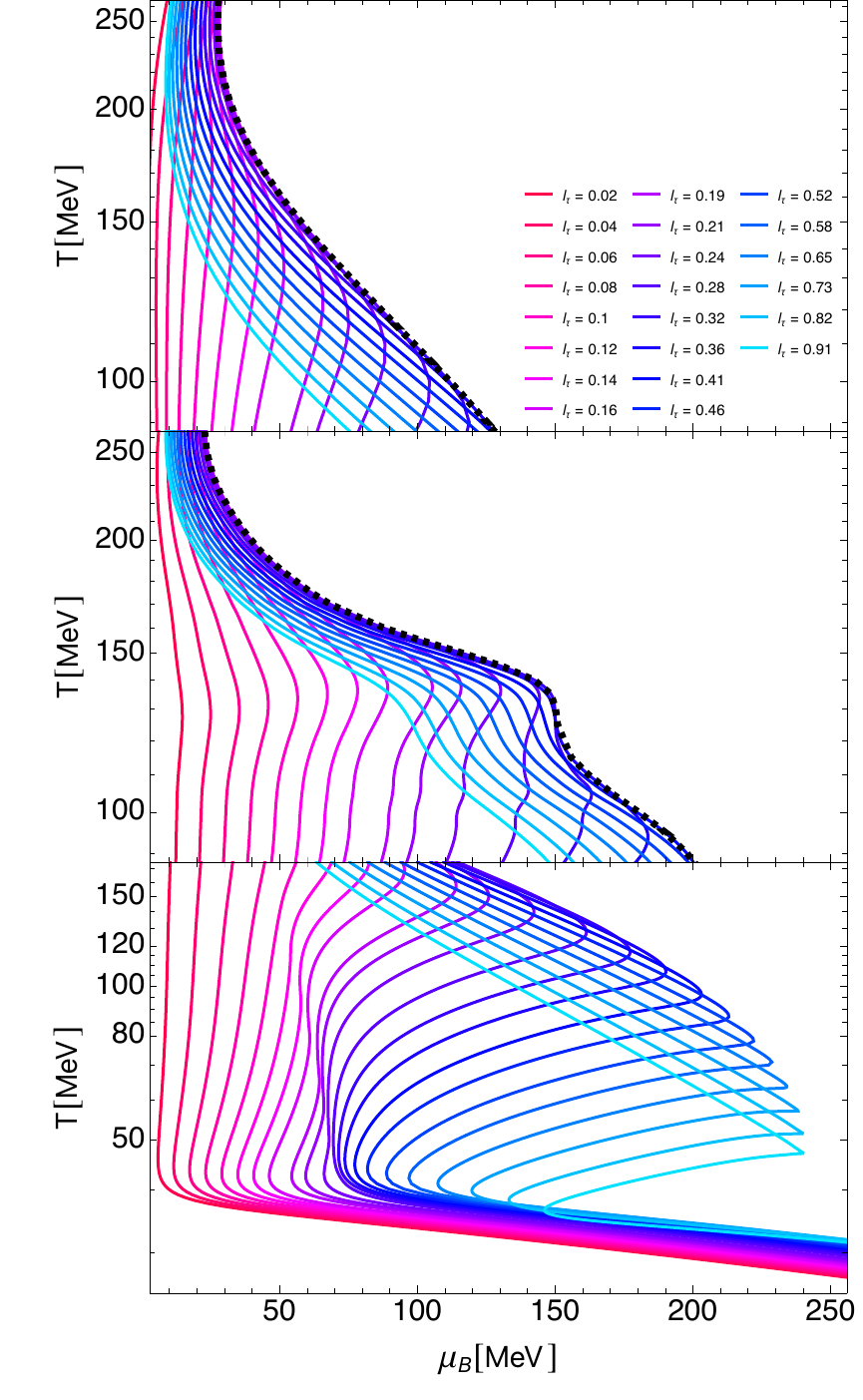}
   \put(52,33){$\mathsf{HRG}$}
    \put(37.67,64){$\mathsf{Lattice-QCD\,\,EoS}$} 
     \put(47.0,95){$\mathsf{Free\,\,QGP}$ }
\end{overpic}
\end{centering}
\caption{Cosmic trajectories in the $T$--$\mu_{\mathrm B}$ plane for $\ell_e \simeq 0$ with $\ell_\mu=-\ell_\tau$ ($\ell_\tau>0$), shown for several values of $\ell_\tau$. Top: free QGP EoS. Middle: lattice QCD-based EoS. Bottom: pure HRG behavior calculated with the Thermal-FIST package \cite{Vovchenko:2019pjl}. The dashed curve on the top and middle panels connects the points where $\mu_\tau = m_\tau$ along each trajectory.}\label{fig:le0cosmic}
\end{figure}

Then, at even lower temperatures, the $\mu^+$ population becomes thermally suppressed as well, and the leptonic charge $Q_\mathrm{lep}\to 0$. 
Charge neutrality is then achieved with small $\mu_\mathrm{Q}$, and the consistent solution returns close to lower $\mu_\mathrm{B}$.

This completes the inversion of the trajectory in the $(T,\mu_\mathrm{B})$ plane.

Using a lattice-based EoS instead of the free QGP one preserves the qualitative picture, namely the existence of a local maximum in $\mu_\mathrm{B}(T)$ for any given cosmic trajectory that emerges when the $\tau$ sector becomes ineffective ($\mu_\tau\simeq m_\tau$).
As we can see in \autoref{fig:le0cosmic}, the main difference between the free QGP and the lattice-based EoS lies in the quantitative location of the local maximum, which, in the case of the lattice-based EoS, is influenced by the transition to the HRG regime at low temperature. The behavior of the pure HRG is shown in the bottom panel of \autoref{fig:le0cosmic}.

In the lattice-based EoS case, the peak appears at \emph{higher} temperatures as $\ell_\tau$ increases. This is in contrast with the free QGP, where a larger $|\ell_\tau|$ tends to move the peak toward \emph{lower} $T$.

At lower temperatures, $T \lesssim 100$ MeV one can use hadron-resonance-gas based equations of state.
We computed the trajectories utilizing the model used in \cite{Vovchenko:2020crk}, which includes pion interactions and the possibility of pion condensation formation.
These calculations are in good agreement with lattice-based EoS results at $T \gtrsim 100$ MeV and show the inversion in $\mu _\mathrm{B}$ at temperatures $50 \lesssim T \lesssim 100$ MeV for $|\ell_\tau| \gtrsim 0.40$, with the maximum $\mu _\mathrm{B}$ never exceeding $250$ MeV.
We also observe that the trajectories for all considered values of asymmetries collapse onto the standard cosmic trajectory in the $T$-$\mu _\mathrm{B}$ plane at $T \lesssim 40$ MeV.

The interacting nature of the lattice-based EoS (and of HRG model) compared to the free quark limit makes a simple qualitative explanation of the shift to lower temperatures difficult because it changes the way charge neutrality is shared between the leptonic and the baryonic sectors.

Nevertheless, our main conclusion does not change: for small total lepton asymmetry $\ell$ allowed by observational constraints, all trajectories are limited in the maximum $\mu_\mathrm{B}$ they can reach. It is worth noting that, when the total lepton asymmetry is fixed to $\ell \sim -10^{-2}$ rather than the $\ell \sim 0$ case shown in \autoref{fig:le0cosmic} (hence $\ell_e \sim 10^{-2}$), the trajectories are almost identical, differing only by a shift of a few MeV toward positive $\mu_\mathrm{B}$.

\begin{figure}
\begin{centering}
\hspace{-1cm}
\includegraphics[width=\columnwidth]{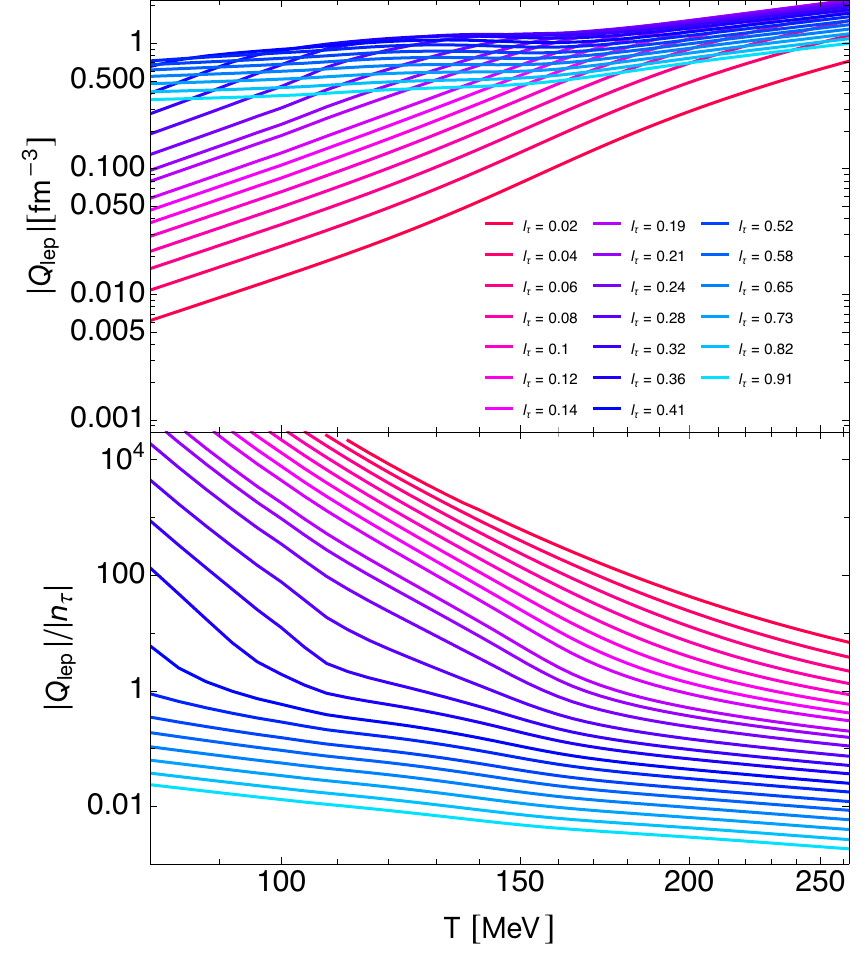}
\end{centering}
\caption{Top: absolute net leptonic charge density $|Q_{\mathrm{lep}}(T)|$ for several $\ell_\tau$ (with $\ell_\mu=-\ell_\tau$) for the lattice QCD-based EoS, where 
$Q_{\mathrm{lep}}(T)\equiv n_{\tau^-}(T)-n_{\mu^+}(T)$. 
Bottom: relative charged-lepton imbalance $\big[n_{\tau^-}(T)-n_{\mu^+}(T)\big]/n_{\tau^-}(T)$ (normalized to the $\tau^-$ density). 
Using the absolute (net) difference, the curves cross; with the relative one, they do not cross and are ordered, indicating that the relative charge imbalance sets the limiting criterion for the trajectories.}\label{fig:le0density}
\end{figure}

When looking at the absolute lepton density contributions to the total charge ($Q_{\rm lep}$), one observes that the curves become ordered according to their relative contributions, as shown in \autoref{fig:le0density}. This behavior indicates that the non-monotonic dependence on $\ell_\tau$ arises from a delicate balance within the charged-lepton sector, which must be compensated by adjustments in the baryonic sector to maintain overall charge neutrality. By contrast, the neutral-lepton (neutrino) sector shows no inversion or non-monotonic ordering. This further indicates that the peculiar behavior arises from the electric-charge neutrality constraint, not from general lepton dynamics.

\textit{Asymmetric case $\ell_\tau \simeq 0$.}--- We now consider the complementary configuration in which the $\tau$-flavor asymmetry is negligible, $\ell_\tau\simeq 0$, while $\ell_e=-\ell_\mu$. In this regime, the $\tau$ sector is suppressed throughout the temperature range of interest, so the charged-lepton budget relevant for charge neutrality is effectively controlled by electrons and muons only:
\begin{equation}
Q_{\rm lep}(T)\simeq\big[n_{e^+}-n_{e^-}\big]+\big[n_{\mu^+}-n_{\mu^-}\big].
\end{equation}
Since $m_e\ll T_\mathrm{QCD}$ and $m_\mu\sim T_\mathrm{QCD}$, both species remain thermally active down to lower temperatures than $\tau$ leptons, and their contributions vary smoothly with $T$.

The absence of an active $\tau$ population removes the heavy-species threshold that, in the $\ell_e\simeq 0$ case, generated a pronounced rebalancing in the leptonic charge and a peak in $\mu_\mathrm{B}(T)$. Here the leptonic sector tends to self-balance over a broader temperature interval, so electric neutrality can be satisfied with only a modest baryonic adjustment. As a result, the single trajectory displays no strong inversion in the $\mu_\mathrm{B}$ direction and has smaller values of $\mu_\mathrm{B}$ at fixed $T$ than in the $\tau$-asymmetric case.

\begin{figure}
\begin{centering}
\hspace{-1cm}

\begin{overpic}[width=1\linewidth,keepaspectratio]{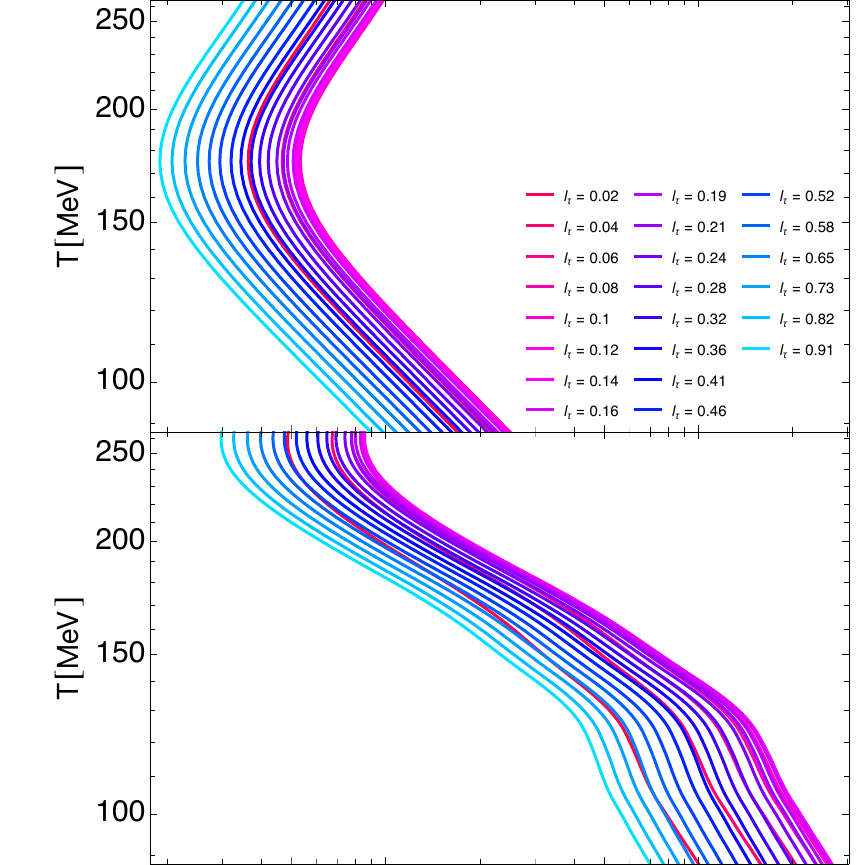}
    \put(58,43){$\mathsf{Lattice-QCD\,\,EoS}$}
     \put(73.2,93){$\mathsf{Free\,\,QGP}$}
\end{overpic}
\end{centering}

\caption{Cosmic trajectories in the $T$--$\mu_{\mathrm B}$ plane for $\ell_\tau \simeq 0$ with $\ell_e=-\ell_\mu$ ($\ell_\mu>0$) , shown for several values of $\ell_\mu$. Top: free QGP EoS. Bottom: lattice QCD-based EoS. Unlike the $\ell_e\simeq 0$ case, there is no sharp rebound: the curves approach an upper limiting trajectory that caps $\mu_\mathrm{B}(T)$ and increasing $|\ell_\mu|$ beyond this does not raise $\mu_\mathrm{B}$ further.}\label{fig:ltau0cosmic}
\end{figure}

Therefore, varying $\ell_\mu$ at fixed $\ell_\tau\simeq 0$ does not produce an inversion in $\mu_\mathrm{B}(T)$, because there is no sharp switch-off of a heavy charged lepton. Instead, as $|\ell_\mu|$ increases, the system approaches a \emph{limiting trajectory} in the $(T,\mu_\mathrm{B})$ plane: the family of trajectories bunches toward an asymptote rather than developing a peak.

As in the previous case, trajectories with larger $|\ell_\mu|$ can lie to the left (smaller $\mu_{\mathrm B}$) of those with smaller $|\ell_\mu|$. This occurs because the relative density difference between the two charged-lepton species (muons and electrons) is reduced, so the leptons themselves contribute more to the charge neutrality (see \autoref{fig:ltau0cosmic}).

\begin{figure}
\begin{centering}
\hspace{-1cm}
\includegraphics[width=\columnwidth]{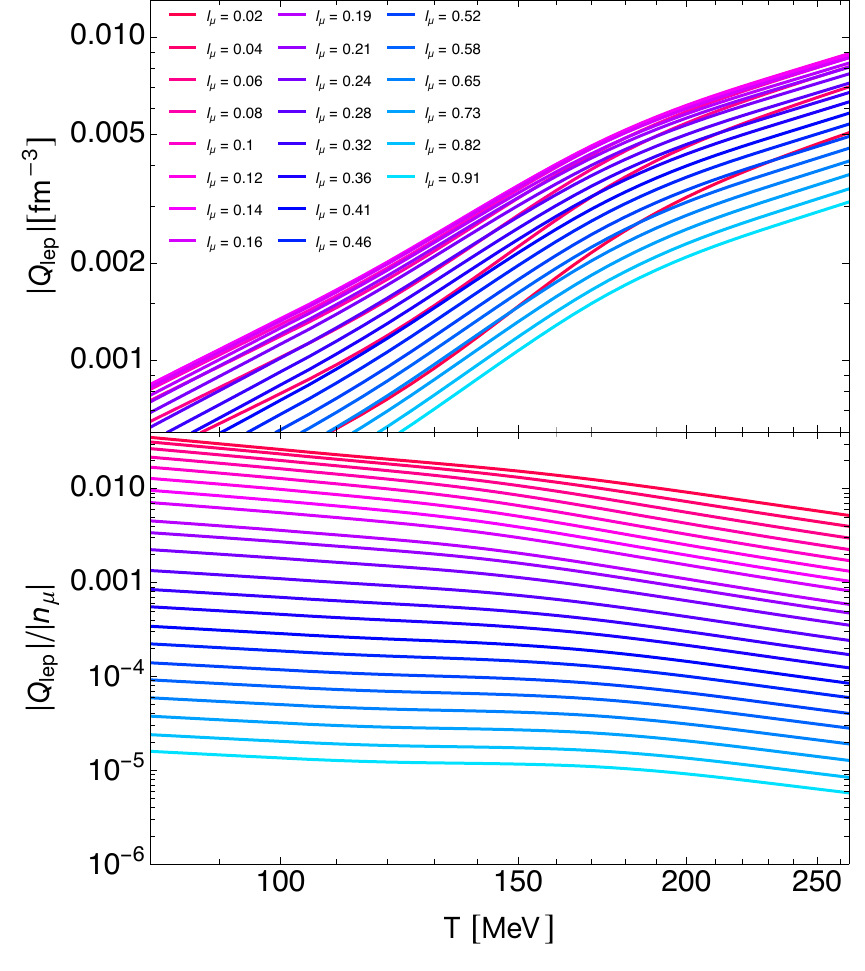}
\end{centering}
\caption{Top: absolute net leptonic charge density $|Q_{\mathrm{lep}}(T)|$ for several $\ell_\mu$ (with $\ell_e=-\ell_\mu$) using the lattice QCD-based EoS, where 
$Q_{\mathrm{lep}}(T)\equiv n_{\mu^-}(T)-n_{e^+}(T)$. 
Bottom: relative charged-lepton imbalance $\big[n_{\mu^-}(T)-n_{e^+}(T)\big]/n_{\mu^-}(T)$ (normalized to the $\mu^-$ density). 
With the absolute (net) difference, the curves reproduce the ordering seen in the $(T,\mu_{\mathrm B})$ trajectories; with the relative difference, the curves are strictly ordered indicating that the relative charge imbalance sets the limiting criterion for the trajectories in this case as well.}\label{fig:ltau0density}
\end{figure}

The ordering of the trajectories can also be characterized by the relative charged-lepton density difference between muons and electrons, which  varies monotonically with $\ell_\mu$ (\autoref{fig:ltau0density}). Notably, as in the $\ell_e\simeq 0$ case, any non-monotonic features in the trajectories originate in the charged-lepton sector; the neutral-leptons (neutrinos) sector shows no such behavior.

\textit{Conclusions.}---
We have revisited the influence of flavor-asymmetric lepton asymmetries on the cosmic trajectory across the QCD epoch, using both a $2+1+1$-flavor lattice-QCD equation of state and a free QGP model, as well the HRG model at low temperatures.  
We showed for the first time that, in the configurations studied, the conserved-charge constraints and the lepton-mass hierarchy impose a robust upper limit on the baryon chemical potential reached along the trajectory.  
While the position of the $\mu_\mathrm{B}(T)$ maximum depends on the chosen EoS, the qualitative behavior is the same: for small total lepton asymmetry ($|\ell|\lesssim10^{-2}$), the trajectory never attains values of $\mu_\mathrm{B}/T$ large enough to cross the first-order QCD region unless we consider that, for a non-zero charge chemical potential $\mu_\mathrm{Q}$, the critical point shifts to low enough $\mu_\mathrm{B}$.

This result contrasts with earlier suggestions that large flavor-asymmetric lepton asymmetries could drive the Universe through a first-order transition when the total lepton asymmetry is negligible. Our analysis shows that the limit in $\mu_\mathrm{B}(T)$ is a consequence of charge neutrality and $\beta$-equilibrium: as the charged leptons become suppressed, the baryon sector compensates, producing a slight, but not arbitrarily large increase in $\mu_\mathrm{B}$.

In particular, the largest values of $\mu _\mathrm{B}(T)$ are observed for the electrophobic scenario ($\ell_e = 0$) for large $\ell_\tau = -\ell_\mu$ but they do not exceed $\mu _\mathrm{B} \sim 250$ MeV.
Present lattice QCD simulations rule out the presence of a first-order phase transition at chemical potentials $\mu _\mathrm{B} < 450$ MeV, at least in the isospin-symmetric scenario \cite{Borsanyi:2025dyp}.

Hence, to reach a first-order transition with the cosmic trajectory, one would need \emph{non-standard} cosmological scenarios. Examples include the “little inflation’’ mechanism \cite{Boeckel:2009ej,Boeckel:2011yj}, Affleck-Dine baryogenesis \cite{Affleck:1984fy}, or other models with early enhanced $\eta_\mathrm{B}$ and/or $\ell$ values. In this way, in principle, the trajectory could be pushed to larger $\mu_\mathrm{B}/T$, where the QCD first-order region resides, and might produce cosmological relics such as density inhomogeneities, strangelets, or primordial black holes \cite{DiClemente:2024lzi,Gouttenoire:2023bqy,Gonin:2025uvc}.  
They may also source a stochastic gravitational-wave background from bubble collisions and turbulence, potentially detectable by pulsar-timing arrays \cite{NANOGrav:2023gor}.

Therefore, within the standard cosmological scenario, small total lepton asymmetries, and flavor-asymmetric configurations cannot trigger a first-order QCD phase transition.  
Nevertheless, these results highlight the importance of exploring non-standard early-Universe conditions to understand the possible links between the QCD phase structure, baryogenesis, and the origin of cosmological relics which are nowadays of a great interest.

\begin{acknowledgements}
This material is based upon work supported by the National Science Foundation under grants No. PHY-2208724, PHY-2116686 and PHY-2514763, and within the framework of the MUSES collaboration, under Grant No. OAC-2103680. This material is also based upon work supported by the U.S. Department of Energy, Office of Science, Office of Nuclear Physics, under Award Numbers DE-SC0022023 and DE-SC0026065, as well as by the National Aeronautics and Space Agency (NASA) under Award Number 80NSSC24K0767.   
\end{acknowledgements}

\bibliography{apssamp}

@misc{Borsanyi:2025dyp,
    author = "Borsanyi, Szabolcs and Fodor, Zoltan and Guenther, Jana N. and Parotto, Paolo and Pasztor, Attila and Ratti, Claudia and Vovchenko, Volodymyr and Wong, Chik Him",
    title = "{Lattice QCD constraints on the critical point from an improved precision equation of state}",
    eprint = "2502.10267",
    archivePrefix = "arXiv",
    primaryClass = "hep-lat",
    month = "2",
    year = "2025"
}

@article{Vovchenko:2019pjl,
    author = "Vovchenko, Volodymyr and Stoecker, Horst",
    title = "{Thermal-FIST: A package for heavy-ion collisions and hadronic equation of state}",
    eprint = "1901.05249",
    archivePrefix = "arXiv",
    primaryClass = "nucl-th",
    doi = "10.1016/j.cpc.2019.06.024",
    journal = "Comput. Phys. Commun.",
    volume = "244",
    pages = "295--310",
    year = "2019"
}

@article{Vovchenko:2020crk,
    author = {Vovchenko, Volodymyr and Brandt, Bastian B. and Cuteri, Francesca and Endr{\H{o}}di, Gergely and Hajkarim, Fazlollah and Schaffner-Bielich, J{\"u}rgen},
    title = "{Pion Condensation in the Early Universe at Nonvanishing Lepton Flavor Asymmetry and Its Gravitational Wave Signatures}",
    eprint = "2009.02309",
    archivePrefix = "arXiv",
    primaryClass = "hep-ph",
    doi = "10.1103/PhysRevLett.126.012701",
    journal = "Phys. Rev. Lett.",
    volume = "126",
    number = "1",
    pages = "012701",
    year = "2021"
}

@article{Abuali:2025tbd,
    author = "Abuali, Ahmed and Bors{\'a}nyi, Szabolcs and Fodor, Zolt{\'a}n and Jahan, Johannes and Kahangirwe, Micheal and Parotto, Paolo and P{\'a}sztor, Attila and Ratti, Claudia and Shah, Hitansh and Trabulsi, Seth A.",
    title = "{New 4D lattice QCD equation of state: Extended density coverage from a generalized T' expansion}",
    eprint = "2504.01881",
    archivePrefix = "arXiv",
    primaryClass = "hep-lat",
    doi = "10.1103/2dmh-26yh",
    journal = "Phys. Rev. D",
    volume = "112",
    number = "5",
    pages = "054502",
    year = "2025"
}

@article{Witten:1984rs,
    author = "Witten, Edward",
    title = "{Cosmic Separation of Phases}",
    reportNumber = "PRINT-84-0400 (IAS,PRINCETON)",
    doi = "10.1103/PhysRevD.30.272",
    journal = "Phys. Rev. D",
    volume = "30",
    pages = "272--285",
    year = "1984"
}

@article{DiClemente:2024lzi,
    author = "Di Clemente, Francesco and Casolino, Marco and Drago, Alessandro and Lattanzi, Massimiliano and Ratti, Claudia",
    title = "{Strange quark matter as dark matter: 40~yr later, a reappraisal}",
    eprint = "2404.12094",
    archivePrefix = "arXiv",
    primaryClass = "hep-ph",
    doi = "10.1093/mnras/staf087",
    journal = "Mon. Not. Roy. Astron. Soc.",
    volume = "537",
    number = "2",
    pages = "1056--1069",
    year = "2025"
}

@article{Affleck:1984fy,
    author = "Affleck, Ian and Dine, Michael",
    title = "{A New Mechanism for Baryogenesis}",
    reportNumber = "Print-84-0574 (PRINCETON)",
    doi = "10.1016/0550-3213(85)90021-5",
    journal = "Nucl. Phys. B",
    volume = "249",
    pages = "361--380",
    year = "1985"
}

@article{Gouttenoire:2023bqy,
    author = "Gouttenoire, Yann",
    title = "{First-Order Phase Transition Interpretation of Pulsar Timing Array Signal Is Consistent with Solar-Mass Black Holes}",
    eprint = "2307.04239",
    archivePrefix = "arXiv",
    primaryClass = "hep-ph",
    doi = "10.1103/PhysRevLett.131.171404",
    journal = "Phys. Rev. Lett.",
    volume = "131",
    number = "17",
    pages = "171404",
    year = "2023"
}

@article{NANOGrav:2023gor,
    author = "Agazie, Gabriella and others",
    collaboration = "NANOGrav",
    title = "{The NANOGrav 15 yr Data Set: Evidence for a Gravitational-wave Background}",
    eprint = "2306.16213",
    archivePrefix = "arXiv",
    primaryClass = "astro-ph.HE",
    doi = "10.3847/2041-8213/acdac6",
    journal = "Astrophys. J. Lett.",
    volume = "951",
    number = "1",
    pages = "L8",
    year = "2023"
}

@article{Musco:2023dak,
    author = "Musco, Ilia and Jedamzik, Karsten and Young, Sam",
    title = "{Primordial black hole formation during the QCD phase transition: Threshold, mass distribution, and abundance}",
    eprint = "2303.07980",
    archivePrefix = "arXiv",
    primaryClass = "astro-ph.CO",
    doi = "10.1103/PhysRevD.109.083506",
    journal = "Phys. Rev. D",
    volume = "109",
    number = "8",
    pages = "083506",
    year = "2024"
}

@article{Gonin:2025uvc,
    author = {Gonin, Ma{\"e}l and Hasinger, G{\"u}nther and Blaschke, David and Ivanytskyi, Oleksii and R{\"o}pke, Gerd},
    title = "{Primordial black-hole formation and heavy r-process element synthesis from the cosmological QCD transition}",
    eprint = "2505.05463",
    archivePrefix = "arXiv",
    primaryClass = "hep-ph",
    doi = "10.1140/epja/s10050-025-01639-w",
    journal = "Eur. Phys. J. A",
    volume = "61",
    number = "7",
    pages = "170",
    year = "2025"
}

@article{Gao:2024fhm,
    author = "Gao, Fei and Harz, Julia and Hati, Chandan and Lu, Yi and Oldengott, Isabel M. and White, Graham",
    title = "{Baryogenesis and first-order QCD transition with gravitational waves from a large lepton asymmetry}",
    eprint = "2407.17549",
    archivePrefix = "arXiv",
    primaryClass = "hep-ph",
    reportNumber = "MITP-24-060",
    doi = "10.1007/JHEP06(2025)247",
    journal = "JHEP",
    volume = "06",
    pages = "247",
    year = "2025"
}

@article{Borsanyi:2020fev,
    author = "Borsanyi, Szabolcs and Fodor, Zoltan and Guenther, Jana N. and Kara, Ruben and Katz, Sandor D. and Parotto, Paolo and Pasztor, Attila and Ratti, Claudia and Szabo, Kalman K.",
    title = "{QCD Crossover at Finite Chemical Potential from Lattice Simulations}",
    eprint = "2002.02821",
    archivePrefix = "arXiv",
    primaryClass = "hep-lat",
    doi = "10.1103/PhysRevLett.125.052001",
    journal = "Phys. Rev. Lett.",
    volume = "125",
    number = "5",
    pages = "052001",
    year = "2020"
}

@misc{Formaggio:2025nde,
    author = "Formaggio, Lorenzo and Di Clemente, Francesco and Yadav, Geetika and Drago, Alessandro and Ratti, Claudia",
    title = "{Cosmic Trajectories calculation with state of the art lattice QCD equation of state}",
    eprint = "2508.00094",
    archivePrefix = "arXiv",
    primaryClass = "astro-ph.CO",
    month = "7",
    year = "2025"
}

@article{PhysRevLett.121.201302,
  title = {Cosmic QCD Epoch at Nonvanishing Lepton Asymmetry},
  author = {Wygas, Mandy M. and Oldengott, Isabel M. and B\"odeker, Dietrich and Schwarz, Dominik J.},
  journal = {Phys. Rev. Lett.},
  volume = {121},
  issue = {20},
  pages = {201302},
  numpages = {6},
  year = {2018},
  month = {Nov},
  publisher = {American Physical Society},
  doi = {10.1103/PhysRevLett.121.201302},
  url = {https://link.aps.org/doi/10.1103/PhysRevLett.121.201302}
}

@article{PhysRevD.105.123533,
  title = {Cosmic QCD transition for large lepton flavor asymmetries},
  author = {Middeldorf-Wygas, Mandy M. and Oldengott, Isabel M. and B\"odeker, Dietrich and Schwarz, Dominik J.},
  journal = {Phys. Rev. D},
  volume = {105},
  issue = {12},
  pages = {123533},
  numpages = {10},
  year = {2022},
  month = {Jun},
  publisher = {American Physical Society},
  doi = {10.1103/PhysRevD.105.123533},
  url = {https://link.aps.org/doi/10.1103/PhysRevD.105.123533}
}

@article{PhysRevLett.128.131301,
  title = {Cosmology Meets Functional QCD: First-Order Cosmic QCD Transition Induced by Large Lepton Asymmetries},
  author = {Gao, Fei and Oldengott, Isabel M.},
  journal = {Phys. Rev. Lett.},
  volume = {128},
  issue = {13},
  pages = {131301},
  numpages = {6},
  year = {2022},
  month = {Mar},
  publisher = {American Physical Society},
  doi = {10.1103/PhysRevLett.128.131301},
  url = {https://link.aps.org/doi/10.1103/PhysRevLett.128.131301}
}

@article{Castorina:2012md,
    author = "Castorina, Emanuele and Franca, Urbano and Lattanzi, Massimiliano and Lesgourgues, Julien and Mangano, Gianpiero and Melchiorri, Alessandro and Pastor, Sergio",
    title = "{Cosmological lepton asymmetry with a nonzero mixing angle $\theta_{13}$}",
    eprint = "1204.2510",
    archivePrefix = "arXiv",
    primaryClass = "astro-ph.CO",
    reportNumber = "CERN-PH-TH-2012-089, IFIC-12-28, LAPTH-018-12",
    doi = "10.1103/PhysRevD.86.023517",
    journal = "Phys. Rev. D",
    volume = "86",
    pages = "023517",
    year = "2012"
}

@article{Zheng2025,
  title = {Quantitative analysis of the gravitational wave spectrum sourced from a first-order chiral phase transition of QCD},
  author = {Zheng, Hui-wen and Gao, Fei and Bian, Ligong and Qin, Si-xue and Liu, Yu-xin},
  journal = {Phys. Rev. D},
  volume = {111},
  issue = {2},
  pages = {L021303},
  numpages = {7},
  year = {2025},
  month = {Jan},
  publisher = {American Physical Society},
  doi = {10.1103/PhysRevD.111.L021303},
  url = {https://link.aps.org/doi/10.1103/PhysRevD.111.L021303}
}

@article{Boeckel:2011yj,
    author = "Boeckel, Tillmann and Schaffner-Bielich, Jurgen",
    title = "{A little inflation at the cosmological QCD phase transition}",
    eprint = "1105.0832",
    archivePrefix = "arXiv",
    primaryClass = "astro-ph.CO",
    doi = "10.1103/PhysRevD.85.103506",
    journal = "Phys. Rev. D",
    volume = "85",
    pages = "103506",
    year = "2012"
}

@article{Shao:2024ygm,
    author = "Shao, Jingdong and Mao, Hong and Huang, Mei",
    title = "{Nanohertz gravitational waves and primordial quark nuggets from dense QCD matter in the early Universe}",
    eprint = "2410.00874",
    archivePrefix = "arXiv",
    primaryClass = "hep-ph",
    doi = "10.1088/1674-1137/adbeeb",
    journal = "Chin. Phys. C",
    volume = "49",
    number = "6",
    pages = "065103",
    year = "2025"
}

@article{Boeckel:2009ej,
    author = "Boeckel, Tillmann and Schaffner-Bielich, Jurgen",
    title = "{A little inflation in the early universe at the QCD phase transition}",
    eprint = "0906.4520",
    archivePrefix = "arXiv",
    primaryClass = "astro-ph.CO",
    doi = "10.1103/PhysRevLett.105.041301",
    journal = "Phys. Rev. Lett.",
    volume = "105",
    pages = "041301",
    year = "2010",
    note = "[Erratum: Phys.Rev.Lett. 106, 069901 (2011)]"
}

@article{Domcke:2025lzg,
    author = "Domcke, Valerie and Escudero, Miguel and Fernandez Navarro, Mario and Sandner, Stefan",
    title = "{Lepton flavor asymmetries: from the early Universe to BBN}",
    eprint = "2502.14960",
    archivePrefix = "arXiv",
    primaryClass = "hep-ph",
    reportNumber = "CERN-TH-2025-010, LA-UR-25-20235",
    doi = "10.1007/JHEP06(2025)137",
    journal = "JHEP",
    volume = "06",
    pages = "137",
    year = "2025"
}

@article{Mangano:2011ip,
    author = "Mangano, Gianpiero and Miele, Gennaro and Pastor, Sergio and Pisanti, Ofelia and Sarikas, Srdjan",
    title = "{Updated BBN bounds on the cosmological lepton asymmetry for non-zero $\theta_{13}$}",
    eprint = "1110.4335",
    archivePrefix = "arXiv",
    primaryClass = "hep-ph",
    reportNumber = "IFIC-11-58, DSF-13-2011",
    doi = "10.1016/j.physletb.2012.01.015",
    journal = "Phys. Lett. B",
    volume = "708",
    pages = "1--5",
    year = "2012"
}

@article{SergioPastor_2010,
doi = {10.1088/1742-6596/203/1/012053},
url = {https://doi.org/10.1088/1742-6596/203/1/012053},
year = {2010},
month = {jan},
publisher = {},
volume = {203},
number = {1},
pages = {012053},
author = {Sergio Pastor},
title = {Relic density of neutrinos with primordial asymmetries},
journal = {Journal of Physics: Conference Series}
}

@article{Barenboim_2017,
   title={Resurrection of large lepton number asymmetries from neutrino flavor oscillations},
   volume={95},
   ISSN={2470-0029},
   url={http://dx.doi.org/10.1103/PhysRevD.95.043506},
   DOI={10.1103/physrevd.95.043506},
   number={4},
   journal={Physical Review D},
   publisher={American Physical Society (APS)},
   author={Barenboim, Gabriela and Kinney, William H. and Park, Wan-Il},
   year={2017},
   month=feb }

@article{Oldengott_2017,
   title={Improved constraints on lepton asymmetry from the cosmic microwave background},
   volume={119},
   ISSN={1286-4854},
   url={http://dx.doi.org/10.1209/0295-5075/119/29001},
   DOI={10.1209/0295-5075/119/29001},
   number={2},
   journal={EPL (Europhysics Letters)},
   publisher={IOP Publishing},
   author={Oldengott, Isabel M. and Schwarz, Dominik J.},
   year={2017},
   month=jul, pages={29001} }

@article{Lattanzi:2024hnq,
    author = "Lattanzi, Massimiliano and Moretti, Mauro",
    title = "{Lepton Asymmetries in Cosmology}",
    doi = "10.3390/sym16121657",
    journal = "Symmetry",
    volume = "16",
    number = "12",
    pages = "1657",
    year = "2024"
}
\end{document}